\documentclass{article}
\usepackage[margin = 1in]{geometry}
\usepackage{natbib}
\RequirePackage[OT1]{fontenc}
\RequirePackage{amsthm,amsmath}

\usepackage{graphicx, amssymb, mathrsfs,amsfonts,url, bm}
\usepackage{hyperref, cleveref, autonum}
\usepackage{color, soul}
\usepackage{enumerate}
\usepackage{multirow}
\usepackage{rotating} 
\usepackage{subfigure} 
\usepackage{cancel}
\usepackage{changes}
\usepackage{booktabs}
\usepackage{graphicx} 
\usepackage{xparse}
\usepackage[boxed]{algorithm2e}
\usepackage{color}
\definecolor{mydarkblue}{rgb}{0,0.08,0.45}
\usepackage{hyperref}
\hypersetup{
	colorlinks=true, 
	linktoc=all,     
	linkcolor=mydarkblue,  
	anchorcolor=mydarkblue,
	citecolor=mydarkblue,
}

\usepackage[T1]{fontenc}
\usepackage[font=small,labelfont=bf,tableposition=top]{caption}
\DeclareCaptionLabelFormat{andtable}{#1~#2  \&  \tablename~\thetable}

\theoremstyle{plain}
\theoremstyle{remark}         
\theoremstyle{definition} 

\NewDocumentCommand{\emphbf}{O{G}}{\emph{\textbf{#1}}}

\usepackage{amsmath}
\usepackage{amsthm} 
\usepackage{graphicx, amssymb, mathrsfs,amsfonts,url, bm}

\author{Jun Lu, \,\,\,\,\, Minhui Wu\\
\texttt{July 8, 2022}
}
\date{}

\title{A note on VIX for postprocessing quantitative strategies}

\begin{document}
\maketitle 

\abstract{
In this note, we introduce how to use Volatility Index (VIX) for postprocessing quantitative strategies so as to increase the Sharpe ratio and reduce trading risks. The signal from this procedure is an indicator of trading or not on a daily basis.
Finally, we analyze this procedure on SH510300 and SH510050 assets. The strategies are evaluated by measurements of Sharpe ratio, max drawdown, and Calmar ratio. However, there is always a risk of loss in trading. The results from the tests are just examples of how the method works; no claim is made on the suggestion of real market positions.
}

\section{Introduction}
The Volatility Index (VIX) (a.k.a., the cash VIX) was introduced by the Chicago Board Options Exchagne (CBOE) in 1993 \citep{sinclair2013volatility}. It was designed to be a benchmark index for equity market volatility. 
To be more concrete, the VIX can be calculated from a weighted strip of options:
$$
\begin{aligned}
\sigma^2_{VIX} &= \frac{2}{T} \sum_{i=1}^{N} \frac{\Delta x_i}{x_i^2} \exp\{rT\}V(x_i)
-\frac{1}{T}\left(\frac{F}{x_0}-1\right)^2;\\
F&= x_0 + \exp\{rT\}(C_0-P_0) ; \\
\Delta x_i &= \frac{x_{i+1}+ x_{i-1}}{2},
\end{aligned}
$$
where $r$ is the risk-free rate, $T$ is the expiration time of the option, $F$ is the forward price of the index, $x_0$ is the strike price immediately below the forward price, $x_i$ is the strike of the $i$-th out-of-the-money (OTM) option, and $V$ is the midprice of the corresponding option. One can refer to \citet{carr2006tale, sinclair2013volatility} for more details on the calculation of VIX values. 
\citet{simon2014vix} shows the futures can be predicted by looking at the VIX. 
That is,  if the futures are trading below the
cash VIX they will tend to rise, and if the futures are trading over the
cash VIX the futures will tend to fall.
The basic basis trade is 
\begin{itemize}
\item If the front futures are above the cash and the expected daily convergence
is greater than 0.1 VIX points, we short a future;
\item If the front futures are below the cash and the expected daily convergence
is greater than 0.1 VIX points, we buy a future.
\end{itemize}
However, not following the basis exactly, we will show how to use the trend of VIX to reduce the risk of a quantitative strategy. We evaluate the strategy in terms of Sharpe ratio (SR) \citep{sharpe1994sharpe, sharpe1966mutual},  Calmar ratio \citep{magdon2004maximum}, and max drawdown (MDD) \citep{magdon2004maximum}; where the larger the better for the former two measurements, and the smaller the better for the last one.

\section{Postprocessing via Effective Ratio of VIX}\label{sec:pp_vix}

\begin{figure}[h]
\centering
\includegraphics[width=0.95\linewidth]{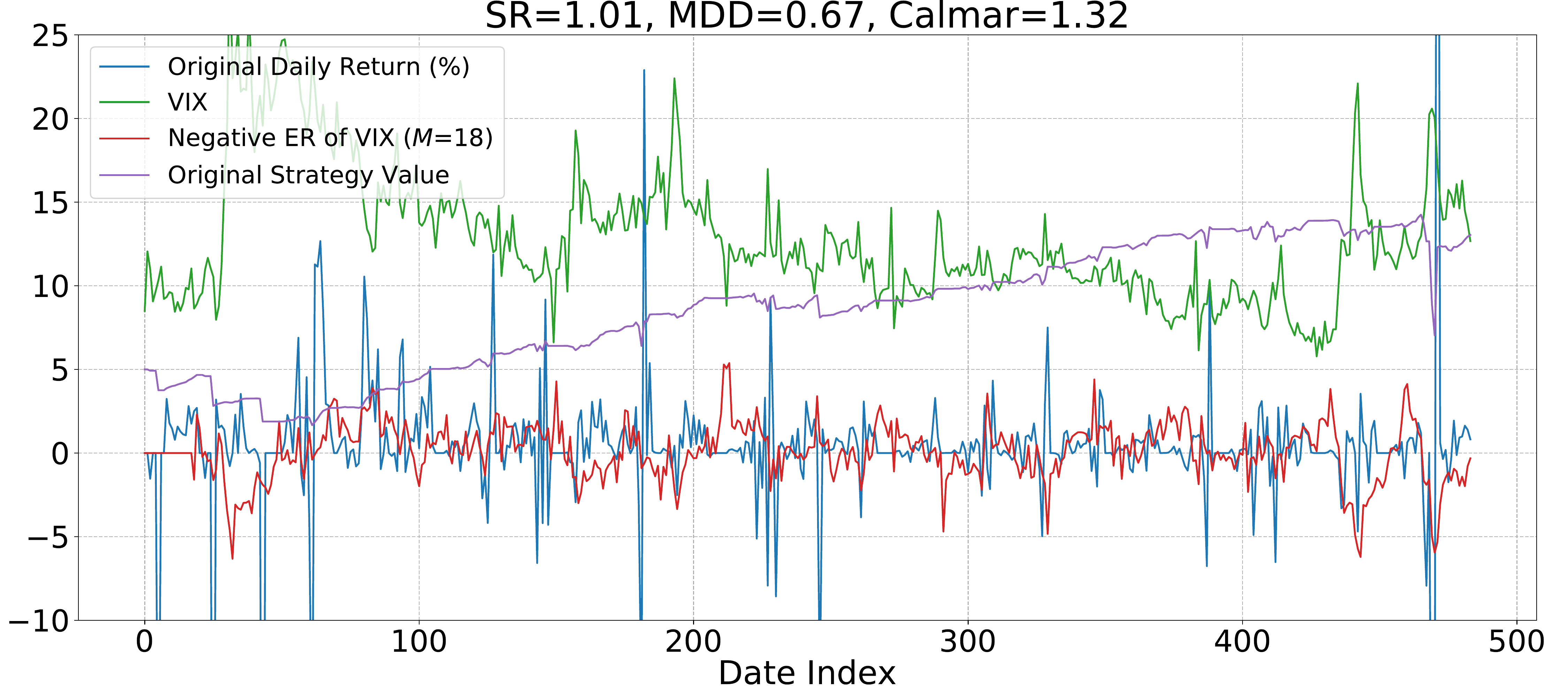}
\caption{An overview of a quantitative strategy on the \textbf{SH510300} dataset, showing the daily return, strategy value, VIX value of the asset, and the negative ER values of the VIX series. The Sharpe ratio (SR), max drawdown (MDD), and Calmar ratio of the strategy are 1.01, 0.67, and 1.32 respectively. The VIX values are divided by 10, and the ER values are multiplied by 10 for clarity.}
\label{fig:510300_overview}
\end{figure}
Figure~\ref{fig:510300_overview} shows a daily-based quantitative strategy where we trade and rebalance every day on SH510300 asset \footnote{The data is from Sina Finance: \url{https://money.finance.sina.com.cn}}.
The \textit{Original Daily Return} represents the daily return (\%) of the strategy, the \textit{Original Strategy Value} indicates the portfolio value of the strategy where initialize the portfolio with a value of 5 for clarity; the strategy here is used only for discussing the postprocessing procedure and we shall not reveal the details. One can plugin into classic quantitative strategies, e.g., \citet{lu2022exploring}.
The \textit{VIX} in the figure indicates the VIX value of the asset.
We can observe there is always a big drawdown (the daily return is approaching -10\% in some cases) when the VIX value is rising up.
Motivated by this finding, we can augment the strategy via the trend of the VIX value. In practice, the trend of the VIX value can be modeled using the \textit{effective ratio} (ER).

\paragraph{Effective ratio}
\citet{kaufman2013trading, kaufman1995smarter} suggested replacing the "weight" variable in the exponential moving average (EMA) formula with a constant based on the \textit{efficiency ratio}. And the ER is shown to provide promising results for financial forecasting via classic quantitative strategies where the ER of the closing price is calculated to decide the trend of the asset \citep{lu2022exploring}. This indicator is designed to measure the \textit{strength of a trend}, defined within a range from \textbf{-1.0 to +1.0} where the larger magnitude indicates a larger upward or downward trend. Instead of calculating the ER of the closing price, we want to calculate the ER of the VIX series. Given the window size $M$, it is calculated with a simple formula:
\begin{equation}\label{equation:effectiveratio-p}
\begin{aligned}
e_t(M)  &= \frac{s_t}{n_t}= \frac{v_{t-1} - v_{t-1-M}}{\sum_{i=1}^{M} |v_{t-i} - v_{t-1-i}|}= \frac{\text{Total VIX change for a period}}{\text{Sum of absolute VIX change for each bar}},
\end{aligned}
\end{equation}
where $e_t(M)$ is the ER at time $t$ and $v_t$ the VIX value at time $t$. We carefully notice that the ER at time $t$ is based on $v_{t-1}, v_{t-2}, \ldots, v_{t-1-M}$ to avoid a forward bias since we do not know $v_t$ at time $t$.
At a strong trend (i.e., the input VIX is moving in a certain direction, up or down) the ER will tend to 1 in absolute value; if there is no directed movement, it will be a little more than 0. The effective ratio is actually a versatile tool, it can also be utilized to find the minimum or maximum point in a time period for the time series, in which case identifying the minimal or maximal point can be noisy in some sense. Recently, \citet{lu2022adasmooth} shows the ER can be applied to stochastic optimization to select the smoothing constant in an adaptive fashion making the optimizer converge faster and continue to reduce the training loss for machine learning tasks; and \citet{lu2022reducing} shows the ER can be applied in augmenting the volatility prediction for reducing overestimating and underestimating the volatility in the peak and trough areas.
\begin{figure}[h!]
	\centering
	\includegraphics[width=0.6\linewidth]{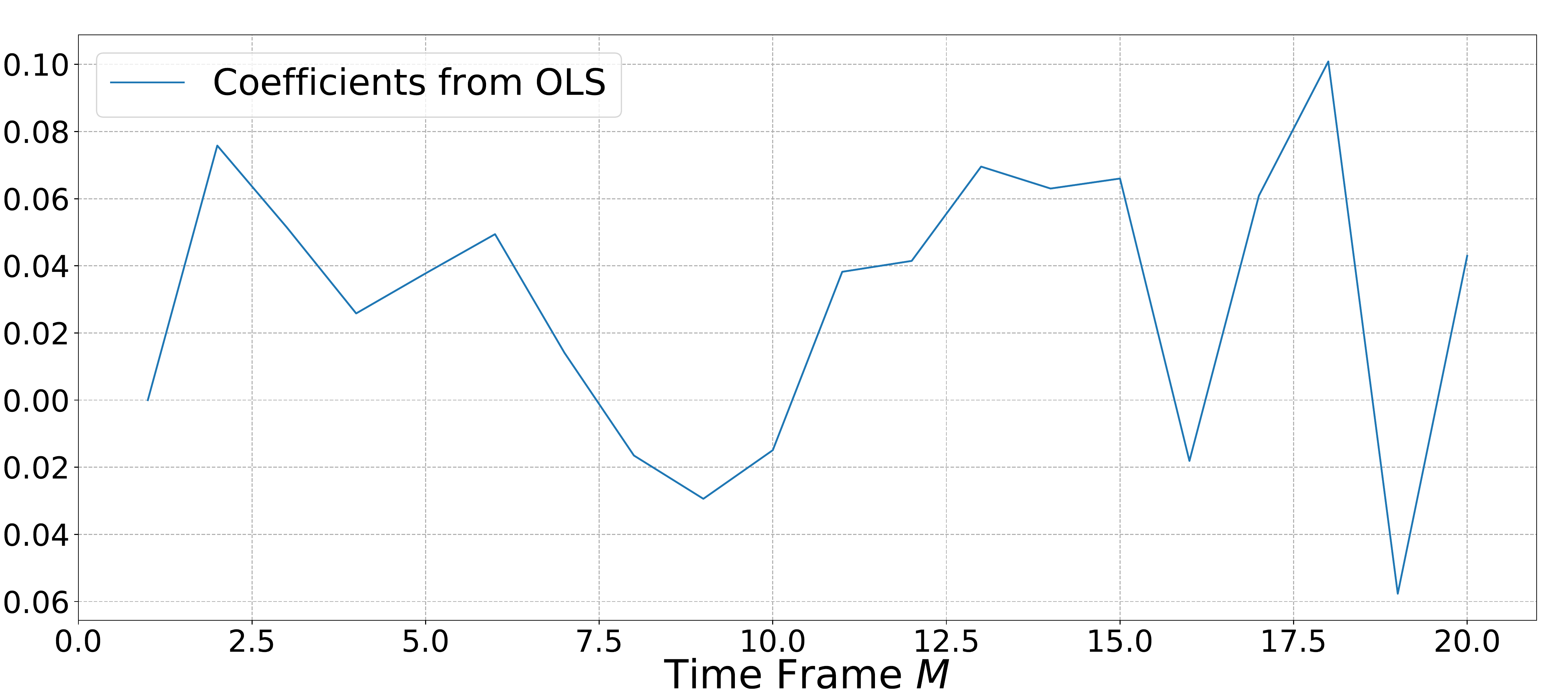}
	\caption{Coefficient values of negative ER and daily return series on \textbf{SH510300} for different time frame $M$.}
	\label{fig:510300_coefficients}
\end{figure}

\begin{figure}[h!]
\centering
\includegraphics[width=0.95\linewidth]{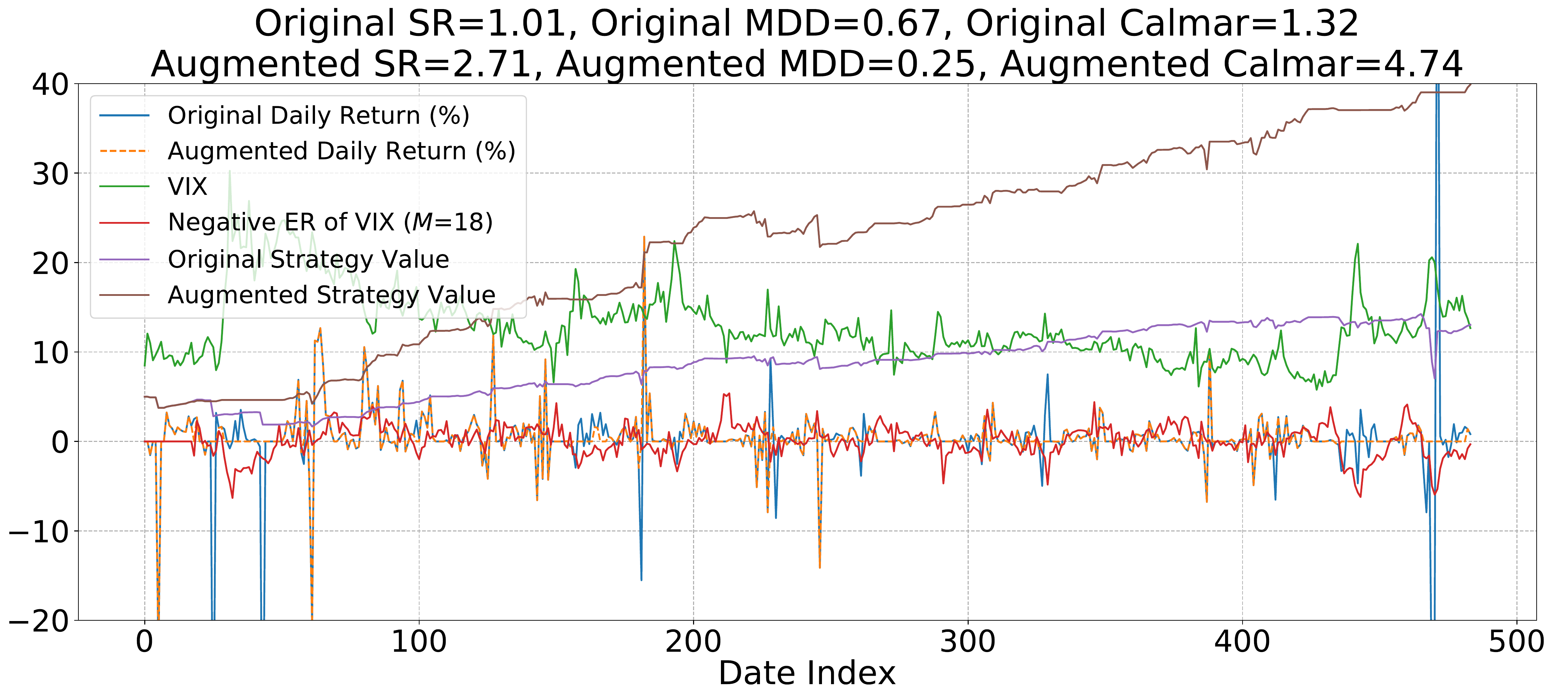}
\caption{An overview of the augmented strategy on the \textbf{SH510300} dataset, showing the daily return, strategy value, VIX value of the asset, and the negative ER values of the VIX series. The Sharpe ratio, max drawdown, and Calmar ratio of the original strategy are 1.01, 0.67, and 1.32 respectively; and of the augmented strategy are 2.71, 0.25, and 4.74 respectively.}
\label{fig:510300_strategy}
\end{figure}

\paragraph{Postprocessing} Having the ER series of VIX values and we recap that there is a big drawdown when the VIX value is rising up. This indicates there might be a positive correlation between the return series and the \textbf{negative} ER of VIX. The only parameter left we need to tune is the time frame $M$ in Eq.~\eqref{equation:effectiveratio-p}.
In practice, the relationship between the negative ER and the return series can be represented by the coefficient value from ordinary least squares (OLS) optimization.
Figure~\ref{fig:510300_coefficients} shows the coefficients of negative ER and daily return series for different time frame $M$ on the SH510300 asset. The time frame varies from 1 trading day to 20 trading days (approximately 1 month). When $M=18$, the coefficient achieves the largest value, approximately 0.1. 
The red curve in Figure~\ref{fig:510300_overview} shows the negative ER value when $M=18$. The curve indicates that the daily return is extremely poor when the negative ER is smaller than -1 (though the value can be tuned further via a loop operation); and we shall simply stop trading on that day. Figure~\ref{fig:510300_strategy} further shows the augmented strategy via this VIX-based postprocessing procedure. We note that the Sharpe ratio increases from 1.01 to 2.71; the MDD increases from 0.67 to 0.25; and the Calmar ratio increases from 1.32 to 4.74. 
In practice, a Sharpe ratio larger than 1 is thought of as a decent strategy; and when the ratio is larger than 2, the strategy is considered extremely excellent.
This simple VIX-based postprocessing procedure can help reduce risk in which case we avoid trading on the days when the volatility is changing rapidly. Though, in some cases, we lose some trading opportunities; however, trading risk can be reduced.

\begin{figure}[h!]
	\centering
	\includegraphics[width=0.6\linewidth]{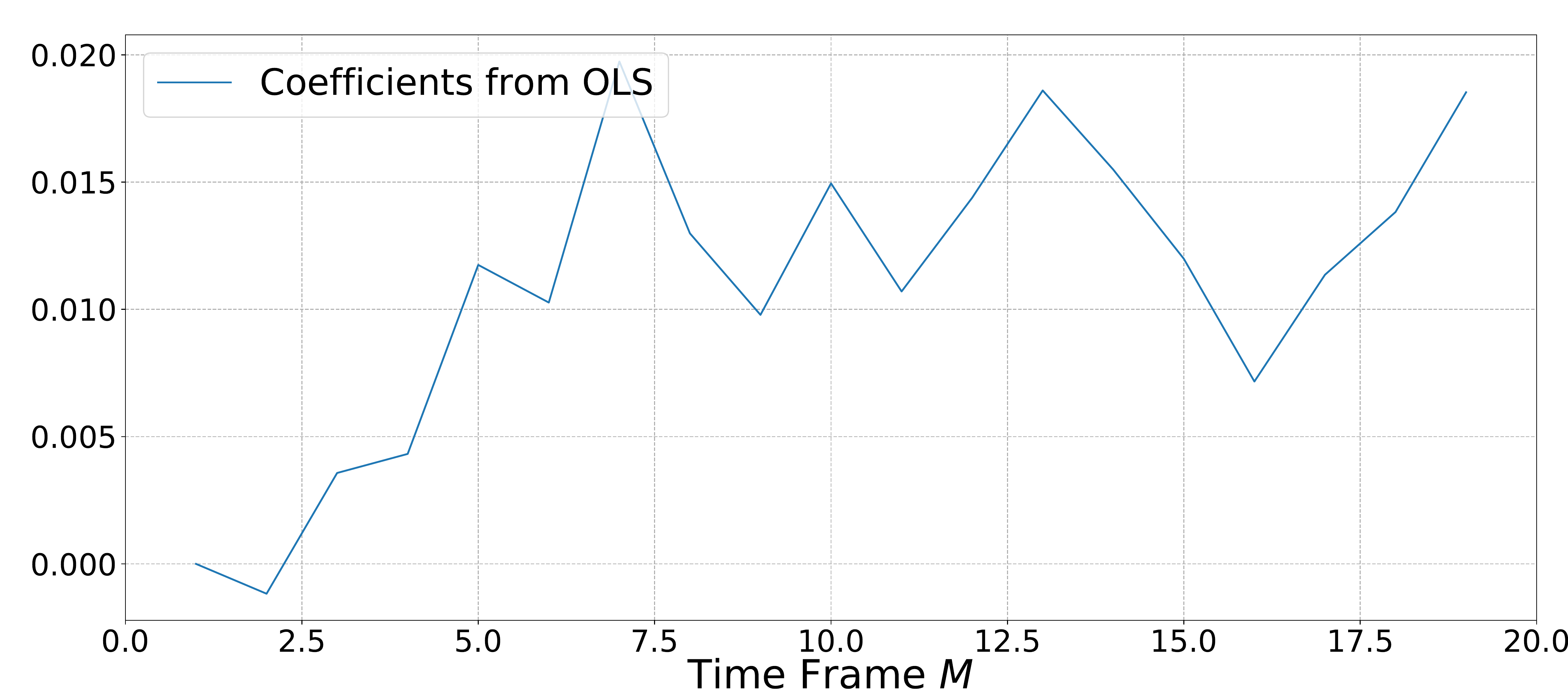}
	\caption{Coefficient values of negative ER and daily return series on \textbf{SH510050} for different time frame $M$.}
	\label{fig:510050_coefficients}
\end{figure}

\begin{figure}[h!]
	\centering
	\includegraphics[width=0.95\linewidth]{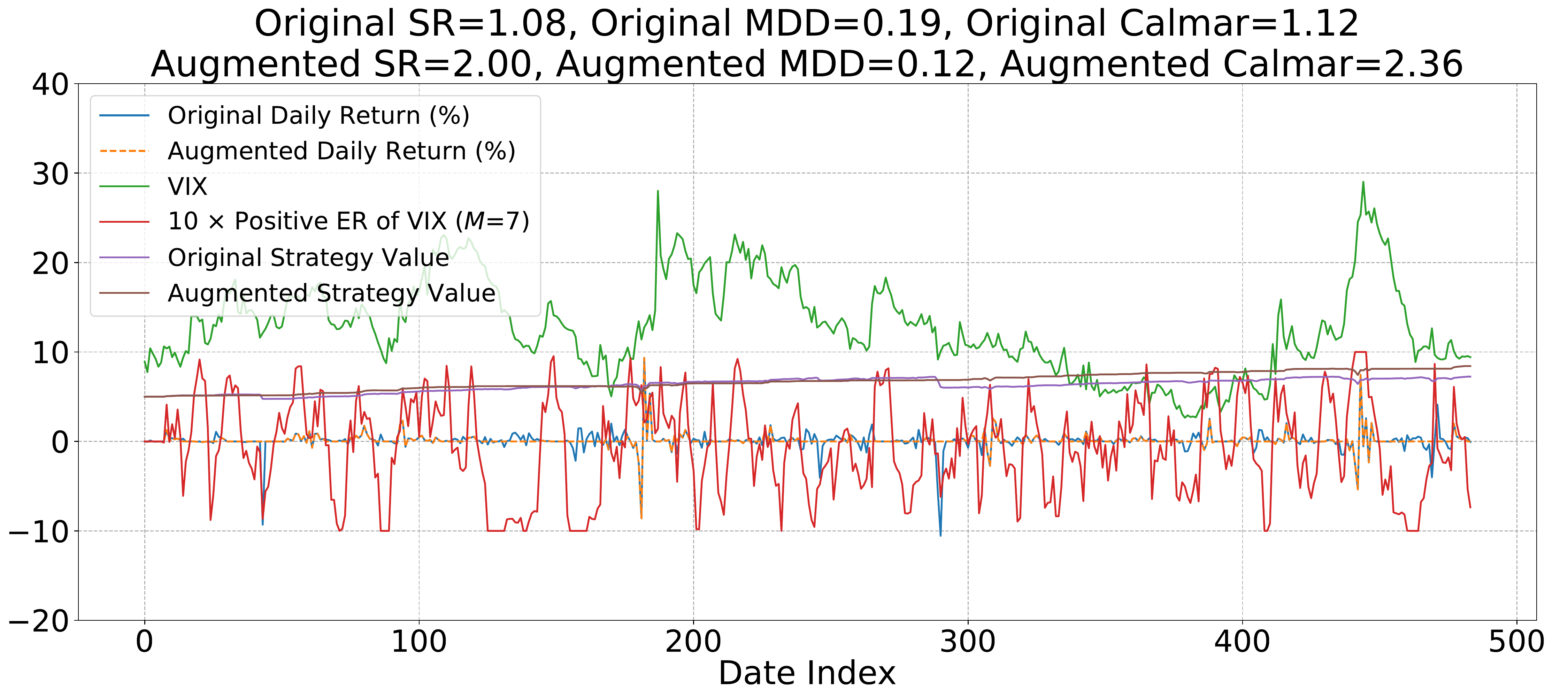}
	\caption{An overview of the augmented strategy on the \textbf{SH510050} dataset, showing the daily return, strategy value, VIX value of the asset, and the \textbf{positive} ER values of the VIX series. The Sharpe ratio, max drawdown, Calmar ratio of the original strategy are 1.08, 0.19, and 1.12 respectively; and of the augmented strategy are 2.00, 0.12, and 2.36 respectively.}
	\label{fig:510050_strategy}
\end{figure}

The postprocessing procedure can be applied on other datasets, e.g., SH510050. 
However, not being dogmatic, we observe there is a positive correlation between the \textbf{positive} ER values and the daily returns. Figure~\ref{fig:510050_coefficients} shows the coefficients between \textbf{positive} ER value and daily return series where the coefficient achieves the largest value when $M=7$.
Similarly, we can augment the strategy on the SH510050 asset as shown in Figure~\ref{fig:510050_strategy} where the  Sharpe ratio is increased from 1.08 to 2.00; the MDD is decreased from 0.19 to 0.12; and the Calmar ratio is increased from 1.12 to 2.36. Though the end portfolio value is not increased significantly (7.27 to 8.53), the risk is reduced to a large extent.

\section*{Risk Disclaimer (for Live-trading)}
There is always a risk of loss in trading. All trading strategies are used at your own risk.
The postprocessing procedure is just for study purposes only. There is no guarantee the exact same algorithm can work on other strategies.

\clearpage
\bibliography{bibliography}
\bibliographystyle{myplainnat}
\end{document}